\documentstyle[12pt]{article}

\def\be{\begin{equation}}
\def\ee{\end{equation}}
\def\bea{\begin{eqnarray}}
\def\eea{\end{eqnarray}}

\begin{document}
\vspace*{2cm}
\begin{center}
{\large\bf D-BRANE DUALITIES 
AS CANONICAL TRANSFORMATIONS}
\vskip 1.5 cm
{\bf Y. Lozano}\footnote{\tt Y.Lozano@fys.ruu.nl}
\vskip 0.05cm
Inst. for Theoretical Physics, Utrecht University,\\
3508 TA Utrecht, The Netherlands
\end{center}

\date{ }
\setcounter{page}{0} \pagestyle{empty}
\thispagestyle{empty}
\vskip 3 cm
\begin{abstract}
We show that the SL(2,R) duality symmetry of type IIB
superstring theory 
can be formulated as the canonical
transformation interchanging momenta and magnetic degrees of
freedom associated to the abelian world-volume gauge field
of the D-3-brane.
D-strings are shown to be connected under the corresponding
transformation in the world-sheet to the $(m,n)$ family of
string solutions of type IIB supergravity
constructed by Schwarz. For the type IIA
superstring the D-2-brane is mapped
under the three-dimensional world-volume
electric-magnetic duality to the dimensional reduction
of the membrane of M-theory.
\end{abstract}
\vfill
\begin{flushleft}
THU-97/03  \\
hep-th/9701186\\
January 1997
\end{flushleft}
\newpage\pagestyle{plain}

\def\theequation{\thesection . \arabic{equation}}

\section{Introduction}
\setcounter{equation}{0}

Recent results in the literature show
certain equivalences between effective actions of
p-branes
of the same or different string (or M-) theories. 
For D-branes \cite{DLP} a relation 
between the D-string and 
the SL(2,Z) multiplet of string solutions of
type IIB supergravity \cite{S} has been proved 
\cite{Sc,DAS,T}, as
well as the self-duality of the D-3-brane under SL(2,Z) 
transformations of
the space-time backgrounds \cite{T,R,GG}.
In both cases the equivalence goes through under certain
transformations in the world-sheet of the D-string 
or the four-dimensional world-volume of the 
D-3-brane\footnote{In \cite{GG} it was shown at the 
level of the equations of motion.}. 
In \cite{T} the general world-volume transformation
of which the above are particular cases was identified and 
referred to as vector duality. 
This vector duality symmetry was interpreted
as the world-volume transformation responsible for the 
strong-weak coupling duality of type IIB \cite{HT,W1,S,S2}, 
in analogy with the 
world-sheet mapping $d\rightarrow *d$ underlying 
T-duality transformations
of the space-time background fields \cite{reviews}.

For the type IIA superstring vector duality is also interesting.
The p-branes of type IIA supergravity have an
alternative interpretation in eleven dimensions which
can be used to deduce its $D=10$ Lorentz covariant world-volume
action \cite{DHIS,To1,To}. 
In the last of the previous references this was done for the 
D-membrane and for the D-4-brane (in this case up to
quadratic order in the gauge field strength) using the fact
that the former can be interpreted as the dimensional
reduction of the eleven dimensional membrane \cite{BST}
and the latter as the double dimensional reduction of the eleven
dimensional 5-brane \cite{G}. For the membrane
in order to prove the complete equivalence between the
dimensionally reduced theory and
the DBI action a non-trivial
transformation involving the abelian world-volume gauge field of 
the D-brane
was shown to be required \cite{Sc,DAS}. Tseytlin pointed out that
this transformation was just the vector
duality symmetry in the three dimensional 
world-volume \cite{T}. 
This observation is in agreement 
with the conjecture that M-theory compactified on a circle
is the strong coupling limit of type IIA \cite{To1,W1}.
This is the reason why a strong-weak coupling duality 
transformation is needed
to connect the type IIA D-2-brane with the membrane of 
M-theory.

In this letter we show that vector duality can be realized
in the phase space of D-branes as the canonical transformation 
interchanging the momenta and the magnetic degrees of freedom
associated to the abelian world-volume
gauge field,
i.e. the same transformation responsible for the SL(2,Z) 
symmetry of electromagnetism\footnote{
SL(2,Z) transformations in non-linear electrodynamics and in
particular of four-dimensional
DBI actions have been studied in \cite{GR}.}.
In $d$ dimensions this transformation maps $r$-forms onto
$(d-r-2)$-forms.

We start in section 2 considering the D-3-brane of type IIB.
In the four-dimensional world-volume the dual of the abelian
gauge field is also a 1-form, and we in fact
show that under the four-dimensional world-volume
canonical transformation a D-3-brane
is obtained in SL(2,R)-transformed backgrounds. 
The SL(2,R) self-duality of the D-3-brane has been shown
previously in \cite{T,R,GG}. This was an
expected result \cite{S} due to the non-existence of
multiplets of 3-brane solutions in type IIB \cite{HS} 
nor bound states of D-3-branes \cite{W2}. 
For the D-string, considered in section 3,
$n$ D-strings are shown to be equivalent under 
the two-dimensional canonical transformation
to the $(m,n)$ family
of string solutions constructed by Schwarz in \cite{S},
supporting the SL(2,Z)
symmetry of type IIB. 
In this case the dual variables are scalars satisfying a 
quantization condition, being this the origin of the $m$
appearing in the dual theory.
The SL(2,Z) multiplet of string solutions,
corresponding to bound states of elementary
and D-strings \cite{W2}, are then shown to be
connected by electric-magnetic world-sheet
duality to D-strings.
  
In section 4 we consider the D-membrane
of type IIA. The result of the canonical transformation is in
this case an eleven dimensional action,  where the
eleventh coordinate has arised as the dual of the abelian 
gauge field in the three-dimensional world-volume. 
This eleven dimensional
action is interpreted as
the dimensional reduction of the membrane of 
M-theory \cite{Sc,DAS,T}.
 
The eleven dimensional 5-brane gives upon double dimensional
reduction the D-4-brane of type IIA \cite{To,BRO,PST}.
As indicated in the last of these references
the equivalence between the bosonic parts of the actions
follows after a vector duality transformation is performed
in the dimensionally reduced action.
In our description this transformation is formulated as
the five-dimensional world-volume canonical transformation
interchanging momenta and magnetic variables.
Details on this construction will be presented elsewhere.

Before considering DBI actions let us recall briefly  
how S-duality in Maxwell's theory is formulated
as a canonical transformation \cite{Y}.
We start by considering the electromagnetic Lagrangian 
with $\theta$-term:
\bea
\label{0.1}
L&=&\frac{1}{8\pi}(\frac{4\pi}{g^2}F_{mn}F^{mn}+
\frac{\theta}{4\pi}
\epsilon^{mnpq}F_{mn}F_{pq}) \nonumber\\
&=&\frac{1}{g^2}(-\vec{E}^2+\vec{B}^2)+\frac{\theta}{4\pi^2}
\vec{E}.\vec{B}
\eea
where $E_\alpha=F_{0\alpha}$, 
$B^\alpha=\frac12\epsilon^{0\alpha\beta\gamma}F_{\beta\gamma}$
and we are taking a Minkowskian space-time.
The canonical momenta are given by:
\bea
\label{0.2}
&&\Pi_0=0\nonumber\\
&&\Pi_\alpha=-\frac{2}{g^2}E_\alpha+\frac{\theta}{4\pi^2}B_\alpha,
\eea
and the Hamiltonian:
\be
\label{0.3}
H=-(\frac{g^2}{4}\vec{\Pi}^2+\frac{g^2\theta}{8\pi^2}
\vec{\Pi}.\vec{B}+
(\frac{g^2{\theta}^2}{64\pi^4}+\frac{1}{g^2})\vec{B}^2)+
\Pi^\alpha\partial_\alpha A_0.
\ee
The primary constraint $\Pi_0=0$ implies the secondary constraint
$\partial_\alpha\Pi^\alpha=0$, therefore the term
$\Pi^\alpha\partial_\alpha A_0$ can be dropped out of the
Hamiltonian but keeping in mind that we are working in the
restricted phase space defined by the two constraints.

It is easy to see that the canonical transformation:
\bea
\label{0.4}
&&{\Pi}^\alpha=
-\frac{1}{2\pi} {\tilde B}^\alpha\nonumber\\
&&{\tilde \Pi}^\alpha
=\frac{1}{2\pi} B^\alpha,
\eea
where ${\tilde F}=d{\tilde A}$, i.e. the interchange between 
electric and magnetic
degrees of freedom\footnote{Note that in the definition of
$\Pi^\alpha$ there is also a contribution from $B^\alpha$ when
$\theta\neq 0$ \cite{Wi}.},
yields the S-dual Hamiltonian with ${\tilde \tau}=-1/\tau$ and
$\tau=\theta/2\pi+4\pi i/g^2$.

The generating functional is given by:
\be
\label{0.5}
G=-\frac{1}{2\pi}\int_{D_4/\partial D_4=M_3}
dA\wedge d{\tilde A}=-\frac{1}{4\pi}\int_{M_3}
d^3x({\tilde A}_\alpha B^\alpha+A_\alpha{\tilde B}^\alpha)
\ee
where $D_4$ is the four-dimensional manifold whose boundary is
the three-dimensional spatial submanifold $M_3$.
In order to show the
complete equivalence between the initial and dual
Hamiltonians we have to prove that they are defined in the same
restricted phase spaces. This is easily seen
by noting that the Bianchi identity component
$\partial_\alpha \,^*F^{0\alpha}=0$ of the initial
theory implies in the dual 
$\partial_\alpha {\tilde \Pi}^\alpha=0$, and ${\tilde \Pi}_0=0$
comes directly from the generating functional above since it
has no ${\tilde A}_0$ dependence. The secondary constraint
$\partial_\alpha\Pi^\alpha=0$ of the original theory implies
in the dual the Bianchi identity component
$\partial_\alpha \,^*{\tilde F}^{0\alpha}=0$.

This same idea can be straightforwardly generalized to
abelian gauge theories of $r$-forms in $d$ dimensions\footnote{Our
conventions are: $\,^*F^{m_1\dots m_{d-r-1}}=
\frac{1}{(r+1)!}\epsilon^{m_1\dots m_d}
F_{m_{d-r}\dots m_d}$; $F\wedge G=\epsilon^{m_1\dots m_d}
F_{m_1\dots m_{r+1}}G_{m_{r+2}\dots m_d}$. In the
particular case $d=2(r+1)$, $r$ odd, a $\theta$-term
can also be introduced in the Lagrangian.}: 
\be
\label{0.6}
S=\int d^dx\frac{1}{2g^2} F\wedge *F.
\ee
The dual theory\footnote{Now
and in the following sections we choose the normalization
of the canonical transformation such that in the 
dual ${\tilde g}=1/g$.}:
\be
\label{0.7}
{\tilde S}=\int d^dx\frac{g^2}{2}
{\tilde F}\wedge *{\tilde F}
\ee
with ${\tilde F}=d{\tilde A}$ and ${\tilde A}$ a
$(d-r-2)$-form, is obtained as the result of the
transformation:
\bea
\label{0.8}
&&\Pi^{\alpha_1\ldots\alpha_r}=\frac{\delta G}{\delta
A_{\alpha_1\ldots\alpha_r}}=
-\,^*{\tilde F}^{0\alpha_1\ldots\alpha_r}\nonumber\\
&&{\tilde \Pi}^{\alpha_1\ldots\alpha_{d-r-2}}=-
\frac{\delta G}{\delta{\tilde A}_{\alpha_1\ldots\alpha_{d-r-2}}}=
(-1)^{dr+d+r-1}\,^*F^{0\alpha_1\ldots\alpha_{d-r-2}},
\eea
with generating functional:
\be
\label{0.9}
G=-\int_{D_d/\partial D_d=M_{d-1}} d^dx 
dA\wedge d{\tilde A}.
\ee
Now the restricted phase space is defined by:
$\Pi^{\alpha_1\dots 0\dots\alpha_r}=0$, 
$\partial_{\alpha_i}\Pi^{\alpha_1\dots\alpha_i\dots\alpha_r}=0$.
The Bianchi identity component
$\partial_{\alpha_i}\,^*F^{0\alpha_1\dots\alpha_i\dots
\alpha_{d-r-2}}=0$ of the initial theory implies
$\partial_{\alpha_i}{\tilde \Pi}^{\alpha_1\dots\alpha_i\dots
\alpha_{d-r-2}}=0$ in the dual, and 
${\tilde \Pi}^{\alpha_1\dots 0\dots\alpha_{d-r-2}}=0$ holds
immediately from the generating functional. Therefore the
original and dual Hamiltonians are defined in the same restricted
phase spaces.

The generating functionals (\ref{0.5}) and (\ref{0.9})
are linear in the original 
and dual variables. Hence we can follow
\cite{GC} and write a relation
between the Hilbert spaces of the original and
dual theories\footnote{Although
still renormalization effects need to be considered.}:
\be
\label{0.10}
\psi_k[{\tilde A}]=N(k)\int {\cal D}A(x^\alpha)
e^{iG[{\tilde A},A(x^\alpha)]}\phi_k[A(x^\alpha)].
\ee
Here $\psi_k$ and $\phi_k$ are eigenfunctions of the respective
Hamiltonians with the same eigenvalue, labelled by $k$,
and $N(k)$ is a normalization factor.
{}From this relation it is possible to derive global properties 
in the dual theory. For instance it is easy to see that if the 
original theory 
satifies a Dirac quantization condition the same holds true for
the dual theory. The equivalence between the original and
the dual partition functions also holds trivially in phase space.

In the next sections we will show that starting 
with D-brane
effective actions we can prove the dualities mentioned in
the introduction as results of the same type
of canonical transformations. 

\section{The D-3-brane of type IIB}
\setcounter{equation}{0}

The effective action of a D-p-brane in RR background fields
contains the usual DBI part\footnote{We are taking a 
Minkowskian signature space-time.} \cite{L}:
\be
\label{1.1}
S_p=\int d^{p+1}x e^{-\phi}\sqrt{-{\rm det}
(G_{mn}+{\cal F}_{mn})},
\ee
where the backgrounds are those induced in the D-brane from  
ten dimensions, ${\cal F}=B+F$ with $F=dA$ and $B$ the
NS-NS two-form; plus a WZ term \cite{LD,Sc,DAS}:
\be
\label{1.2}
S_{WZ}=\int d^{p+1}\theta_0 d^{p+1}x\sum_{r=0}^{p+1}
C_{(r)}(\theta_0)
e^{\frac12 {\cal F}_{mn}\theta^m_0\theta^n_0},
\ee
where $\theta_0$ denote the fermionic zero modes and
$C_{(r)}(\theta_0)=\frac{1}{m_r!}C_{m_1\ldots m_r}
\theta^{m_1}_0\ldots \theta^{m_r}_0$ 
are the induced rank $r$ RR
background fields.
We will focus on the bosonic part of the 
actions, assuming that the fermionic part will be  
fixed by supersymmetry and fermionic kappa symmetry
\cite{kappa}.

The effective action of the D-3-brane in presence of RR 
background fields is then given by:
\be
\label{1.3}
S_3=\int d^4x[e^{-\phi}\sqrt{-{\rm det}
(G_{mn}+{\cal F}_{mn})}+\frac18 \epsilon^{mnpq}(\frac13 C_{mnpq}
+2C_{mn}{\cal F}_{pq}+C{\cal F}_{mn}{\cal F}_{pq})].
\ee
Introducing the dilaton inside the square root
and working with the canonical metric 
$g_{mn}=e^{-\phi/2}G_{mn}$ we can write:
\be
\label{1.4}
e^{-\phi}\sqrt{-{\rm det}(G_{mn}+{\cal F}_{mn})}=
\sqrt{-{\rm det}g}\sqrt{1+\frac12 e^{-\phi}{\cal F}^2+
\frac18 e^{-2\phi}({\cal F}^2)^2-\frac14 e^{-2\phi} 
{\cal F}^4},
\ee
where ${\cal F}^2={\cal F}_{mn}{\cal F}^{mn}$ and
${\cal F}^4={\cal F}_{mn}{\cal F}^{np}{\cal F}_{pq}
{\cal F}^{qm}$. Restricting for simplicity
to a Minkowskian canonical metric
$g_{mn}=\eta_{mn}$ and
introducing electric and magnetic variables:
${\cal E}_\alpha={\cal F}_{0\alpha}$,
${\cal B}^\alpha=\frac12 \epsilon^{0\alpha\beta\gamma}
{\cal F}_{\beta\gamma}$, where the letters ${\cal E}, {\cal B}$
are used to recall their dependence
on the NS-NS two-form, we can write:
\be
\label{1.5}
S_3=\int d^4x [\sqrt{1-e^{-\phi}\vec{{\cal E}}^2
+e^{-\phi}{\vec{\cal B}}^2-
e^{-2\phi}(\vec{{\cal E}}.\vec{{\cal B}})^2}+\vec{C}.
\vec{{\cal B}}
+\vec{D}.\vec{{\cal E}}+C\vec{{\cal E}}.\vec{{\cal B}}+
\frac{1}{24}\epsilon^{mnpq}C_{mnpq}],
\ee
where we have defined $C_\alpha\equiv C_{0\alpha}$ and
$D^\alpha\equiv \,^*C^{0\alpha}$.
The $A_m$-conjugate momenta are given by:
\bea
\label{1.6}
&&\Pi_0=0\nonumber\\
&&\Pi_\alpha=
-\frac{e^{-\phi}}{\sqrt{1-e^{-\phi}\vec{{\cal E}}^2+
e^{-\phi}\vec{{\cal B}}^2-e^{-2\phi}
(\vec{{\cal E}}.\vec{{\cal B}})^2}}
({\cal E}_\alpha+e^{-\phi}(\vec{{\cal E}}.\vec{{\cal B}})
{\cal B}_\alpha)
+D_\alpha+C{\cal B}_\alpha,
\eea
and the Hamiltonian:
\bea
\label{1.7}
&&H_3=-\sqrt{1+e^{-\phi}\vec{{\cal B}}^2+e^\phi (\vec{\Pi}-
(\vec{D}+C\vec{{\cal B}}))^2+\vec{{\cal B}}^2
(\vec{\Pi}-(\vec{D}+C\vec{{\cal B}}))^2-
(\vec{{\cal B}}(\vec{\Pi}-(\vec{D}+
C\vec{{\cal B}})))^2}\nonumber\\
&&-\vec{C}\vec{B}-\Pi^\alpha
B_{0\alpha}-\frac{1}{24}\epsilon^{mnpq}C_{mnpq}+\Pi^\alpha
\partial_\alpha A_0.
\eea
The same canonical transformation responsible for S-duality
in electromagnetism\footnote{Now we have chosen different
normalizations.}:
\bea
\label{1.8}
&&\Pi_\alpha=-{\tilde B}_\alpha\nonumber\\
&&{\tilde \Pi}_\alpha=B_\alpha,
\eea
(in the four-dimensional world-volume the dual theory is
also defined in terms of 1-forms ${\tilde A}$)
can be shown to yield
the effective action of a D-3-brane\footnote{With the same
remarks concerning the constraints.} in the SL(2,R)
transformed backgrounds:
\be
\label{1.9}
e^{-{\tilde \phi}}=\frac{1}{e^{-\phi}+e^{\phi}C^2},\,\,\,
{\tilde C}=-\frac{Ce^\phi}{e^{-\phi}+e^{\phi}C^2},
\ee
or, equivalently:
\be
\label{1.10}
{\tilde \lambda}=-\frac{1}{\lambda},\,\,\,
\lambda\equiv C+ie^{-\phi},
\ee
and
\be
\label{1.11}
{\tilde B}_{mn}=C_{mn},\,\,\,
{\tilde C}_{mn}=-B_{mn},\,\,\,
{\tilde C}_{mnpq}=C_{mnpq}.
\ee
This transformation together with the invariance
of the action under constant shifts of the RR scalar
field generates the whole SL(2,R) invariance, which
is broken to SL(2,Z) due to quantum effects.
The same calculation 
for an arbitrary metric $g_{mn}$ yields
${\tilde g}_{mn}=g_{mn}$, as $g_{mn}$ is the canonical
metric.

Substituting in (\ref{1.8}) the expressions for the 
canonical momenta the non-local change of variables 
in configuration space responsible 
for SL(2,R) transformations is obtained. 
This change of variables is the one induced by the Fourier 
transformation of (\ref{1.3}) with respect to the field 
strength of the abelian world-volume gauge field \cite{T}.
Due to the complicated and
non-linear expressions for the momenta this
cannot be interpreted as a simple generalization of
$d\rightarrow *d$ (or the interchange between
electric and magnetic degrees of freedom).
However in phase space the transformation
is very simple being just the 
definition of S-duality in a four dimensional
abelian gauge theory.
 
We have shown that the D-3-brane is self-dual under
SL(2,R) transformations, with self-duality meaning that
the dual is also a D-3-brane, although defined in the
SL(2,R) transformed backgrounds.
In our formalism it is clear that the reason for this
is the general $A_r\leftrightarrow {\tilde A}_{d-r-2}$
electric-magnetic duality, that particularized to
$d=4$ and $r=1$ yields also a dual 1-form, and 
it is consistent with the non-existence of
multiplets of 3-brane solutions in type IIB
supergravity \cite{HS}.

As was mentioned by Tseytlin
\cite{T} this transformation in the world-volume of the
D-brane can be interpreted as the world-volume
mapping responsible for SL(2,R) transformations of the
space-time backgrounds, in analogy with the $d\rightarrow *d$
world-sheet mapping underlying the T-duality
transformation $R\rightarrow 1/R$ in space-time.
In our construction this symmetry is
understood as the same type of canonical transformation
responsible for S-duality in abelian gauge theories.
Therefore SL(2,Z) can be
viewed as a subgroup of the whole group of simplectic
diffeomorphisms on the world-volume of the 3-brane.
Also, we can implement this
transformation in the path integral
since both the action
and the measure remain invariant in phase space.
This generalizes previous results in the literature
\cite{T,R},
where a saddle point approximation had to be made.

Finally let us mention that four-dimensional
BI actions coupled to an
SL(2,R) invariant four-dimensional bulk were considered
in \cite{GR}
in relation to the S-duality of the heterotic string in
four dimensions \cite{het}. 
In our case the D-3-brane
is coupled to a ten dimensional SL(2,R) invariant bulk 
\cite{SL(2R)} and
the S-duality under consideration is the one of the 
ten dimensional
type IIB superstring.

\section{$(m,n)$ strings in type IIB}
\setcounter{equation}{0}

The same type of analysis above can be made for the D-string.
In this case we find a correspondence with the $(m,n)$
string solutions of type IIB supergravity
constructed by Schwarz in \cite{S}.

We just need to recall the results stated in the introduction
about electric-magnetic duality in $d$ dimensional abelian
gauge theories of arbitrary $r$-forms. Our starting point is the
action describing the bound state of $n$ 
D-strings\footnote{Since $n$ parallel D-p-branes are described
by a $U(1)^n$ gauge theory in the $(p+1)$-dimensional
world-volume \cite{W2} we effectively get a factor of $n$ in
front of the one D-string action.}:
\be
\label{2.1}
S_1=\int d^2x n[e^{-\phi}\sqrt{-{\rm det}(G_{mn}+{\cal F}_{mn})}
+\frac12 \epsilon^{mn}(C_{mn}+C{\cal F}_{mn})].
\ee
In a two-dimensional world-volume there are only
 electric degrees of freedom.
Redefining $g_{mn}=e^{-\phi}G_{mn}$ and taking 
$g_{mn}=\eta_{mn}$ for simplicity we can write:
\be
\label{2.1.1}
S_1=\int d^2x n[\sqrt{1-e^{-2\phi}{\cal E}_1^2}+C_{01}+
C{\cal E}_1].
\ee 
The canonical momenta are given by:
\bea
\label{2.1.2}
&&\Pi_0=0\nonumber\\
&&\Pi_1=-\frac{ne^{-2\phi}}{\sqrt{1-e^{-2\phi}{\cal E}_1^2}}
{\cal E}_1+nC
\eea
and the Hamiltonian:
\be
\label{2.2}
H_1=-\sqrt{n^2+e^{2\phi}(\Pi_1-nC)^2}-\Pi_1B_{01}-nC_{01}
+\Pi_1\partial_1A_0.
\ee
The canonical transformation responsible for electric-magnetic
duality is in this case:
\bea
\label{2.3}
&&\Pi^1=-\,^*{\tilde F}^{01}=-\epsilon^{01}{\tilde \Lambda}
\nonumber\\
&&{\tilde \Pi}=\,^*F^0=\epsilon^{01}A_1,
\eea
with generating functional:
\be
\label{2.5}
G_1=-\oint dx\epsilon^{0\alpha}A_\alpha{\tilde \Lambda}.
\ee
Since we only have electric degrees of freedom the two
transformations in (\ref{2.3}) cannot be independent.
One checks easily that the first one is enough to obtain the
dual theory and the second gives the corresponding
equation of motion.
The secondary constraint $\partial_1\Pi^1=0$, associated to
$\Pi_0=0$, implies that
${\tilde \Lambda}$ must be a function of time, and moreover
from (\ref{0.10}) (note that $G_1$ is 
linear in the original and dual variables)
we obtain that ${\tilde \Lambda}$
must be an integer due to Dirac quantization condition in
the original theory. 
Setting ${\tilde \Lambda}=-m$
we find a dual action:
\be
\label{2.6}
{\tilde S}_1=\int d^2x[\sqrt{n^2+e^{2\phi}(nC-m)^2}
+\frac12\epsilon^{mn}
(nC_{mn}+mB_{mn})].
\ee
For arbitrary metric the result is:
\be
\label{2.7}
{\tilde S}_1=\int d^2x[\sqrt{e^{-2\phi}n^2+(nC-m)^2}
\sqrt{-{\rm det}G_{mn}}+\frac12\epsilon^{mn}(nC_{mn}+
mB_{mn})],
\ee
i.e. the Nambu-Goto action of a fundamental string
with tension $T=\sqrt{e^{-2\phi}n^2+(nC-m)^2}$ and
charges $(m,n)$ with respect to the NS-NS and RR
2-forms. 
The existence of this string multiplet
is required by the SL(2,Z) symmetry of type IIB, since
it is obtained from the elementary type
IIB superstring (the (1,0) string in this notation):
\be
\label{2.8}
S_{(1,0)}=\int d^2x[\sqrt{-{\rm det}G_{mn}}+\frac12
\epsilon^{mn}B_{mn}]
\ee
after an SL(2,Z) transformation
of parameters
\[ \Lambda=\left( \begin{array}{cc}
                  p & q \\
                 -n & m
                  \end{array} \right),
\qquad pm+qn=1. \]
This multiplet was first constructed by Schwarz in \cite{S} as
a set of solutions of type IIB supergravity. There
the argument requiring that $m,n$ were integers and 
relatively prime was Dirac quantization condition plus
stability since if this was
not the case a given
$(m,n)$ string would be at the threshold of decaying into
$p$ $(\frac{m}{p},\frac{n}{p})$ strings, with $p$
the maximum common divisor of $m$ and $n$.
In \cite{W2} Witten showed that these
string solutions correspond to bound states of $m$ 
fundamental strings and $n$ D-strings.
{}From the world-volume point of view its existence
can be proved starting from $n$ D-strings, as shown here and,
previously, in \cite{Sc,DAS,T}.
We have seen that the
transformation that is required is the corresponding 
electric-magnetic duality in a two-dimensional
world-volume, which we
have formulated as a canonical
transformation. The equivalence between the partition
functions also holds straightforwardly in our formalism.

\section{The D-membrane of type IIA}
\setcounter{equation}{0}

Since SL(2,R) is not a symmetry of type IIA supergravity
we don't expect any relation between the D-membrane of type
IIA and other objects in the same theory, under the 
three-dimensional
world-volume transformation.
Instead we expect some conexion with
M-theory (compactified on a circle) since it is
conjectured to be the strong coupling limit of 
ten dimensional type IIA.
As we mentioned in the introduction it is in fact
known that the vector dual of the
D-membrane is the dimensional reduction of the membrane
of M-theory \cite{Sc,DAS,T}. Here we show that this
duality can also be formulated as a canonical transformation
in the three-dimensional world-volume.

We start by considering the effective action of the D-membrane:
\be
\label{3.1}
S_2=\int d^3x[e^{-\phi}\sqrt{-{\rm det}(G_{mn}+{\cal F}_{mn})}
+\frac12\epsilon^{mnp}(\frac13 C_{mnp}+C_m{\cal F}_{np})].
\ee
Redefining $g_{mn}=e^{-2\phi/3}G_{mn}$ and chosing $g_{mn}$ to
be Minkowski for simplicity we can write in electric and
magnetic variables (our conventions are: 
${\cal E}_\alpha={\cal F}_{0\alpha}$, 
${\cal B}=\frac12 \epsilon^{0\alpha\beta}{\cal F}_{\alpha\beta}$):
\be
\label{3.2}
S_2=\int d^3x[\sqrt{1-e^{-4\phi/3}\vec{{\cal E}}^2+e^{-4\phi/3}
{\cal B}^2}+\frac16\epsilon^{mnp}C_{mnp}+C_0{\cal B}+
\vec{D}.\vec{{\cal E}}]
\ee
where 
$D^\alpha\equiv\,^*C^{0\alpha}=\epsilon^{0\alpha\beta}C_\beta$.
The canonical momenta are given by:
\bea
\label{3.2.1}
&&\Pi_0=0\nonumber\\
&&\Pi_\alpha=-\frac{e^{-4\phi/3}{\cal E}_\alpha}{\sqrt{1
-e^{-4\phi/3}\vec{{\cal E}}^2+e^{-4\phi/3}{\cal B}^2}}
+D_\alpha,
\eea
and the Hamiltonian:
\be
\label{3.3}
H_2=-\sqrt{1+e^{4\phi/3}(\vec{\Pi}-\vec{D})^2+e^{-4\phi/3}{\cal B}^2
+{\cal B}^2(\vec{\Pi}-\vec{D})^2}-C_0{\cal B}-\Pi_\alpha B_{0\alpha}
-\frac16\epsilon^{mnp}
C_{mnp}+\Pi_\alpha\partial_\alpha A_0.
\ee
The canonical transformation:
\bea
\label{3.4}
&&\Pi^\alpha=-\,^*{\tilde F}^{0\alpha}=
-\epsilon^{0\alpha\beta}\partial_\beta
{\tilde \Lambda}\nonumber\\
&&{\tilde \Pi}=B
\eea
with generating functional:
\be
\label{3.5}
G_2=-\int_{M_2}d^2x \epsilon^{0\alpha\beta}A_\alpha
\partial_\beta{\tilde \Lambda},
\ee
yields the following dual action
(for arbitrary metric):
\be
\label{3.6}
{\tilde S}_2=\int d^3x[\sqrt{-{\rm det}(e^{-2\phi/3}G_{mn}+
e^{4\phi/3}(\partial_m{\tilde \Lambda}+C_m)(\partial_n
{\tilde \Lambda}+C_n))}+\frac16\epsilon^{mnp}
(C_{mnp}-3\partial_m{\tilde \Lambda}B_{np})].
\ee
This is the Nambu-Goto action of the dimensionally reduced
eleven dimensional
supermembrane \cite{BST}, with:
\bea
\label{3.7}
&&G^{(11)}_{mn}=e^{-2\phi/3}G^{(10)}_{mn}+e^{4\phi/3}(\partial_m
{\tilde \Lambda}+C_m)(\partial_n{\tilde \Lambda}+C_n)\nonumber\\
&&B^{(11)}_{mnp}=C^{(10)}_{mnp}-3\partial_m{\tilde \Lambda}B_{np}.
\eea

In \cite{To} the conexion between the D-membrane and the
dimensionally reduced eleven dimensional supermembrane
was pointed out, and further developed in \cite{Sc,DAS,T}
where it was shown that a given world-volume transformation
on the D-membrane was required. Here we have seen that it
is just the strong-weak coupling transformation
(defined in phase space), needed to 
connect ten dimensional type IIA with M-theory.

\section{Conclusions}
\setcounter{equation}{0}

To summarize, we have seen that certain S-dualities between D-branes
and p-branes of some string (or M-) theories can be formulated
as canonical transformations in the phase space defined by the 
abelian world-volume gauge field of the D-brane and its conjugate 
momentum\footnote{Although the transformation is involutive 
it is more instructive to start from the D-brane
action because then all the dualities can
be formulated in a more unified way.}.
This generalizes previous results in the literature 
\cite{Sc,DAS,T,R,GG} where a saddle point 
approximation is required
in order to prove the equivalence with the corresponding dual
theory. In our description the equivalence holds 
straightforwardly at a quantum mechanical level\footnote{Up to 
renormalization effects, as mentioned previously.}. 

The canonical transformation description provides a unified 
picture
in which duality symmetries in string theories can be
formulated. The same kind of transformation
responsible for S-duality in abelian 
gauge theories is responsible for S-duality in the world-volume
of D-branes.
This transformation is just the interchange
between momenta and magnetic degrees of freedom. For
abelian gauge theories it reduces to 
electric-magnetic exchange
whereas for the D-branes it is quite more complicated
due to the non-trivial dependence of the DBI action on the
electric degrees of freedom.

As mentioned in the introduction the D-4-brane of type IIA
is equivalent to the double dimensional reduction of the
eleven dimensional 5-brane after a vector duality 
transformation is performed in the five-dimensional
world-volume. The dual theory is then defined in terms of
a 2-form. 
The double dimensional ``oxidation'' of this theory
could shed some light on the determination of the
action of the eleven dimensional 5-brane
\cite{To,BRO,PST,Aha,W3}. 

There are more results in the literature which would be
interesting to formulate in the present description.
In \cite{Aha} it is shown that vector duality
in the eleven dimensional membrane gives upon dimensional
reduction T-duality in the ten dimensional type II superstring.
Given that both symmetries have well-defined descriptions
in terms of canonical transformations (see the second of \cite{Y})
it could be interesting
to show their explicit conexion. 

Finally it could also be interesting to study the role of
electric-magnetic duality transformations in the relation
between the p-branes of type IIB and F theories \cite{FT}  
(see for instance \cite{JR} for some recent results).

\subsection*{Acknowledgements}

I would like to thank E. Alvarez, A. Gonz\'alez-Ruiz 
and E. Verlinde for useful discussions.
Work supported by the European Commission TMR
programme ERBFMRX-CT96-0045.


\newpage

\begin{thebibliography}{99}

\medskip

\bibitem{DLP}
J. Dai, R.G. Leigh and J. Polchinski, Mod. Phys. Lett. A4
(1989) 2073;
J. Polchinski, Phys. Rev. Lett. 75 (1995) 4724;
``TASI Lectures on D-branes'', hep-th/9611050.

\bibitem{S}
J.H. Schwarz, Phys. Lett. B360 (1995) 13.

\bibitem{Sc}
C. Schmidhuber, Nucl. Phys. B467 (1996) 146.

\bibitem{DAS}
S. P. de Alwis and K. Sato, Phys. Rev. D53 (1996) 7187.

\bibitem{T}
A.A. Tseytlin, Nucl. Phys. B469 (1996) 51.

\bibitem{R}
S. Ryang, hep-th/9608118.

\bibitem{GG}
M.B. Green and M. Gutperle, Phys. Lett. B377 (1996) 28.

\bibitem{HT}
C.M. Hull and P.K. Townsend, Nucl. Phys. B438 (1995) 109;
C.M. Hull, Phys. Lett. B357 (1995) 545.

\bibitem{W1}
E. Witten, Nucl. Phys. B443 (1995) 85.

\bibitem{S2}
J.H. Schwarz, Phys. Lett. B367 (1996) 97.

\bibitem{reviews}
A. Giveon, M. Porrati and E. Rabinovici, Phys. Rep. 244
(1994) 77; E. Alvarez, L. Alvarez-Gaum\'e and Y. Lozano,
Nucl. Phys. B (Proc. Supp.) 41 (1995) 1.

\bibitem{DHIS}
M.J. Duff, P.S. Howe, T. Inami and K.S. Stelle, Phys. Lett.
B191 (1987) 70;
M.J. Duff and K.S. Stelle, Phys. Lett. B253 (1991) 113;
M.J. Duff and J.X. Lu, Nucl. Phys. B390 (1993) 276;
M.J. Duff, J.T. Liu and R. Minasian, Nucl. Phys. B452 (1995)
261.

\bibitem{To1}
P.K. Townsend, Phys. Lett. B350 (1995) 184.

\bibitem{To}
P.K. Townsend, Phys. Lett. B373 (1996) 68.

\bibitem{BST}
E. Bergshoeff, E. Sezgin and P.K. Townsend, Phys. Lett. B189
(1987) 75.

\bibitem{G}
R. G\"uven, Phys. Lett. B276 (1992) 49.

\bibitem{GR}
G.W. Gibbons and D.A. Rasheed, Nucl. Phys. B454 (1995) 185;
Phys. Lett. B365 (1996) 46.

\bibitem{HS}
G.T. Horowitz and A. Strominger, Nucl. Phys. B360 (1991) 197;
M.J. Duff and J.X. Lu, Phys. Lett. B273 (1991) 409.

\bibitem{W2}
E. Witten, Nucl. Phys. B460 (1996) 335.

\bibitem{BRO}
E. Bergshoeff, M. de Roo and T. Ort\'{\i}n, Phys. Lett. B386
(1996) 85.

\bibitem{PST}
P. Pasti, D. Sorokin and M. Tonin, hep-th/9701037;
M. Aganagic, J. Park, C. Popescu and J.H. Schwarz,
hep-th/97011166.

\bibitem{Y}
Y. Lozano, Phys. Lett. B364 (1995) 19; 
Mod. Phys. Lett. A11 (1996) 2893.

\bibitem{Wi}
E. Witten, Phys. Lett. B86 (1979) 283.

\bibitem{GC}
G.I. Ghandour, Phys. Rev. D35 (1987) 1289;
T. Curtright, in ``Differential Geometrical Methods in 
Theoretical Physics: Physics and Geometry'', eds. L.L. Chau
and W. Nahm, Plenum, New York, (1990) 279; T. Curtright
and G.I. Ghandour, in ``Quantum Field Theory, Statistical
Mechanics, Quantum Groups and Topology'', eds. T.Curtright,
L. Mezincescu and R. Nepomechie, World Scientific, (1992),
hep-th/9503080.

\bibitem{L}
R.G. Leigh, Mod. Phys. Lett. A4 (1989) 2767.

\bibitem{LD}
M. Li, Nucl. Phys. B460 (1996) 351;
C.G. Callan and I.R. Klebanov, Nucl. Phys. B465 (1996) 473;
M. Douglas, hep-th/9512077; 
C.G. Callan, C. Lovelace, C.R. Nappi and S.A. Yost,
Nucl. Phys. B308 (1988) 221.

\bibitem{kappa}
M. Cederwall, A. von Gussich, B.E.W. Nilsson and
A. Westerberg, hep-th/9610148;
M. Cederwall, A. von Gussich, B.E.W. Nilsson, P. Sundell
and A. Westerberg, hep-th/9611159;
E. Bergshoeff and P.K. Townsend, hep-th/9611173;
M. Aganagic, C. Popescu and J.H. Schwarz, hep-th/9612080.

\bibitem{het}
A. Font, L.E. Iba\~nez, D. L\"ust and F. Quevedo,
Phys. Lett. B249 (1990) 35;
A. Sen, Int. J. Mod. Phys. A9 (1994) 3707. 

\bibitem{SL(2R)}
J.H. Schwarz, Nucl. Phys. B226 (1983) 269;
P.S. Howe and P.C. West, Nucl. Phys. B238 (1984) 181;
E. Bergshoeff, C.M. Hull and T. Ort\'{\i}n,
Nucl. Phys. B451 (1995) 547.

\bibitem{Aha}
O. Aharoni, Nucl. Phys. B476 (1996) 470.

\bibitem{W3}
E. Witten, hep-th/9610234;
M. Perry and J.H. Schwarz, hep-th/9611065;
J.H. Schwarz, hep-th/9701008.

\bibitem{FT}
C. Vafa, Nucl. Phys. B469 (1996) 403.

\bibitem{JR}
D.P. Jatkar and S.K. Rama, Phys. Lett. B388 (1996) 283.



\end{thebibliography}
\end{document}